\documentclass[twocolumn, pra, superscriptaddress]{revtex4-1}
\usepackage{amssymb}
\usepackage{amsmath}
\usepackage{epsfig}
\usepackage{color}
\usepackage{graphics, graphicx}
\usepackage{bbold}
\usepackage{psfrag}
\usepackage{mathcomp}
\usepackage{subfigure}
\usepackage{verbatim}
\usepackage{color}
\usepackage[colorlinks,citecolor=blue]{hyperref}
\def\cp#1{\mathbf{#1}}

\begin{document}
\title{Effective scattering and Efimov physics in the presence of two-body dissipation}
\author{Lihong Zhou}
\affiliation{Beijing National Laboratory for Condensed Matter Physics, Institute of Physics, Chinese Academy of Sciences, Beijing 100190, China}
\author{Xiaoling Cui}
\email{xlcui@iphy.ac.cn}
\affiliation{Beijing National Laboratory for Condensed Matter Physics, Institute of Physics, Chinese Academy of Sciences, Beijing 100190, China}
\affiliation{Songshan Lake Materials Laboratory , Dongguan, Guangdong 523808, China}
\date{\today}

\begin{abstract}
Two-body dissipation usually gives rise to a complex interaction. Here, we study the effect of two-body dissipation on few-body physics, including the fundamental two-body effective scattering and the three-body Efimov physics. By employing a two-channel model that incorporates the decay of closed-channel molecules (generating the two-body dissipation), we explicitly relate the real and imaginary part of the inverse scattering length ($a_s^{-1}$) to closed-channel detuning and decay rate. In particular, we show that
the imaginary part of $a_s^{-1}$ is given by the product of the molecule decay rate and the effective range. Such complex scattering length is found to generate an additional imaginary Coulomb potential when three atoms come close to each other, thereby suppressing the formation of trimer bound states and modifying the conventional discrete scaling in Efimov physics. 
\end{abstract}

\maketitle

\section{Introduction}

Dissipation is ubiquitous in nature, while its origin can be associated with different decay processes. The two-body dissipation describes a typical class of decay process where two particles are simultaneously lost from the system due to their pair-wise interaction. Therefore, such dissipation is usually modeled by a complex interaction. In cold atoms system, the two-body dissipation can be realized in magnetically or optically induced Feshbach resonance via the spontaneous or laser-induced decay of intermediate molecule state\cite{Fedichev, Bohn, Lett, Grimm, Takahashi_ofr1, Takahashi_ofr2, Duerr, Zhang, Chin}.  
For atoms in lattices, the two-body dissipation can be effectively tuned by  the lattice depth\cite{Durr, Takahashi1, Takahashi2,Sponselee} in experiments, and it has been found that the strong inelastic collisions can actually inhibit particle losses\cite{Durr, Takahashi2} and more favor the Mott insulating phase instead of bosonic  superfluid\cite{Takahashi1,Takahashi2}. Theoretically,  it has been shown that the two-body dissipation can influence the magnetic correlation\cite{Ueda1, Ueda2} and dynamical properties\cite{Pan,Ripoll,Rossini} of lattice bosons, induce spin-charge separation\cite{Ueda4} and lead to intriguing fermion superfluidity including the dissipation-induced phase transitions\cite{Das, Yamamoto} and superfluid reentrance\cite{Ueda3, Iskin}.

While there have been quite a number of studies on the many-body consequence of two-body dissipation, little is known to the few-body physics modified by it.  For a dilute atomic gas, the few-body physics serves as a fundamental building block for the study of many-body phenomena, since it provides the basic effective interaction model for low-energy particles\cite{Stringari_RMP,Chin_RMP} and determines the practical few-body loss in realistic experiment\cite{Braaten,Efimov_review2,Efimov_review3}. In this work, we will focus on the few-body physics affected by two-body dissipation, including the two-body effective scattering and the three-body Efimov physics.

Considering the origin of two-body dissipation in realistic cold atomic system\cite{Lett, Grimm, Takahashi_ofr1, Takahashi_ofr2, Duerr, Zhang, Chin}, here we employ a two-channel model that incorporates the decay of closed-channel molecules by an imaginary molecule energy (sharing the same spirit as in Refs.\cite{Fedichev, Bohn}). The molecule decay rate characterizes  the strength of two-body dissipation. After the renormalization, we explicitly relate the real and imaginary parts of the inverse s-wave scattering length ($a_s^{-1}$) to the molecule detuning and decay rate. In particular, we show that  the imaginary part of $a_s^{-1}$ is given by the product of the molecule decay rate and the effective range. In the three-body sector, the complex scattering length generates an additional imaginary Coulomb potential when three particles come close to each other. Such potential is found to suppress the formation of trimer bound states, and lead to a generalized discrete scaling law for Efimov physics. We have demonstrated these effects 
using both the Born-Oppenheimer approximation and the hyper-spherical coordinate approach.

The rest of this paper is organized as follows. In section \ref{sec_II}, we introduce the two-channel model and study the effective scattering theory under two-body dissipation. Sections \ref{sec_III} and \ref{sec_IV} respectively contribute to the effect of two-body dissipation to two-body bound state and Efimov physics in three-body sector.  
Finally we summarize our work in section \ref{sec_V}.

\section{Effective scattering under two-body dissipation} \label{sec_II}

We write down the following two-channel model:
\begin{eqnarray}
H&=&\sum_{{\cp q}\sigma} \epsilon_{\cp q} a^{\dag}_{{\cp q}\sigma}a_{{\cp q}\sigma} + \sum_{\cp Q} ({\cal E}_{\cp Q}+\delta-i\Gamma) d^{\dag}_{{\cp Q}}d_{{\cp Q}} \nonumber\\
&&+ \frac{g}{\sqrt{V}} \sum_{{\cp Q}{\cp q}} \left(d^{\dag}_{\cp Q} a_{{\cp Q}-{\cp q},\uparrow} a_{{\cp Q}+{\cp q},\downarrow} + h.c.\right).
\end{eqnarray}
Here $a^{\dag}_{{\cp q}\sigma}$ creates an open-channel atom with spin-$\sigma$ ($\sigma=\uparrow,\downarrow$), momentum ${\cp q}$ and energy $\epsilon_{\cp q}={\cp q}^2/(2m)$;  $d^{\dag}_{{\cp Q}}$ creates a closed-channel molecule with momentum ${\cp Q}$ and kinetic energy ${\cal E}_{\cp Q}={\cp Q}^2/(4m)$; $\delta$ and $\Gamma$  denote, respectively, the molecule detuning and decay rate. Here $\Gamma$ characterizes the strength of two-body dissipation and gives rise to an effective complex interaction for open-channel atoms. In this work we take $\hbar$ as unity.

We now derive the scattering matrix of two colliding open-channel atoms($\uparrow$ and $\downarrow$) with relative momentum ${\cp k}$ and energy $E=k^2/m\ (k\equiv|{\cp k}|)$. First we write down the Lippman-Schwinger equation $T=U+UG_0T$, where $G_0=(E-H_0)^{-1}$ is the non-interacting Green function, and the bare interaction $U$ can be obtained via the virtual scattering to closed-channel molecules, which gives $U=g^2/(E-(\delta-i\Gamma))$. Finally we have:
\begin{equation}
\frac{1}{T(E)} =\frac{E-(\delta-i\Gamma)}{g^2} - \frac{1}{V} \sum_{\cp q} \frac{1}{E-2\epsilon_{\cp q}}. \label{T_1}
\end{equation}
By expressing $T(E)$ in terms of the s-wave scattering length $a_s$ and the effective range $r_0$:
\begin{equation}
\frac{1}{T(E)} = \frac{m}{4\pi} \left(\frac{1}{a_s} -\frac{1}{2}r_0 k^2 +i k \right), \label{T_2}
\end{equation}
we then have
\begin{align}
&  \frac{m}{4\pi a_s}=\frac{-\delta+i\Gamma}{g^2} + \frac{1}{V} \sum_{\cp q} \frac{1}{2\epsilon_{\cp q}}; \label{a_s} \\
& r_0=-\frac{8\pi}{m^2g^2}.
\end{align}
Therefore, the real and imaginary parts of $1/a_s$ read:
\begin{align}
&  {\rm Re}(\frac{1}{a_s})=\frac{4\pi}{m} \left( \frac{-\delta}{g^2} + \frac{1}{V} \sum_{\cp q} \frac{1}{2\epsilon_{\cp q}}\right); \label{Re} \\
&  {\rm Im}(\frac{1}{a_s})= \frac{4\pi}{mg^2} \Gamma= \frac{1}{2}m|r_0|\Gamma. \label{Im}
\end{align}
One can see that the two-body dissipation ($\Gamma\neq0$) will not affect the real part of $1/a_s$, but will induce a finite imaginary part of $1/a_s$. This says, the real and imaginary parts of $1/a_s$ can be tuned {\it independently} by the detuning $\delta$ and decay rate $\Gamma$ of closed-channel molecules. In comparison, the real or imaginary part of $a_s$ itself will reply on both $\delta$ and $\Gamma$, but not single of them.
Moreover, Eq.\eqref{Im} tells an important result that the imaginary part of $1/a_s$ is given by the product of $\Gamma$ and effective range $r_0$. Therefore, the effect of two-body dissipation is more pronounced in a narrow Feshbach resonance with large $r_0$, since it will help to produce a larger ${\rm Im}(1/a_s)$.

Note that in the above two-channel model, we have neglected the background scattering of open-channel atoms. In the presence of a finite background scattering length $a_{\rm bg}$ and by replacing $a_s$ with $a_s-a_{\rm bg}$, our results can well match the form of scattering length as in Refs.\cite{Fedichev, Bohn}. 

In literature, there have been extensive studies on the few- and many-body properties for cold atoms with a finite effective range in the absence of two-body dissipation\cite{Blume,Bolda,Petrov, Bruun, Gurarie,Pricoupenko, Dieckmann, Ho, Ohara}. Here,  in order to highlight the effect of ${\rm Im}(1/a_s)$ and simplify the calculation, we omit the effective range term in Eq.(\ref{T_2}), such that the interaction strength is solely determined by the scattering length $a_s$. In the absence of two-body dissipation, this approximation is generally valid for low-energy system near resonance with $|r_0|\ll |a_s|$. 
When the two-body dissipation is present, $1/a_s$ has a finite imaginary part (Eq.(\ref{Im})), and therefore the condition additionally requires $r_0 k^2\ll {\rm Im}(1/a_s)$, i.e., the energy considered should be well below the dissipation strength $\Gamma$. Otherwise, one should take into account the effective range term in Eq.(\ref{T_2}).

 In the following, we will study the two-body and three-body bound states in the presence of two-body dissipation, i.e., with a complex $a_s$. The binding energies of these bound states are defined with respect to the scattering threshold, i.e., $E_{\rm th}=0$ when the two- or three-particles are infinitely apart from each other. Therefore, we only show the bound state solution with a negative real part of the binding energy $({\rm Re}(E_b)<0)$.

\section{Two-body bound state} \label{sec_III}

The two-body binding energy $E_b$ is determined by $T(E_b)=\infty$, which gives $E_b=-1/(ma_s^2)$ and therefore
\begin{align}
&{\rm Re}(E_b)=-\frac{1}{m} \left({\rm Re}^2(\frac{1}{a_s}) - {\rm Im}^2(\frac{1}{a_s})\right);\\
&{\rm Im}(E_b)= -\frac{2}{m} {\rm Re}(\frac{1}{a_s}){\rm Im}(\frac{1}{a_s}).
\end{align}
One can see that a two-body bound state (with ${\rm Re}(E_b)<0$) emerges when ${\rm Re}(\frac{1}{a_s})>{\rm Im}(\frac{1}{a_s})>0$, which is a more stringent requirement than the case without two-body dissipation ($1/a_s>0$). The bound state is associated with a finite life-time $\tau\sim (2{\rm Re}(1/a_s){\rm Im}(1/a_s)/m)^{-1}$.

\begin{figure}[h]
\includegraphics[width=7.5cm]{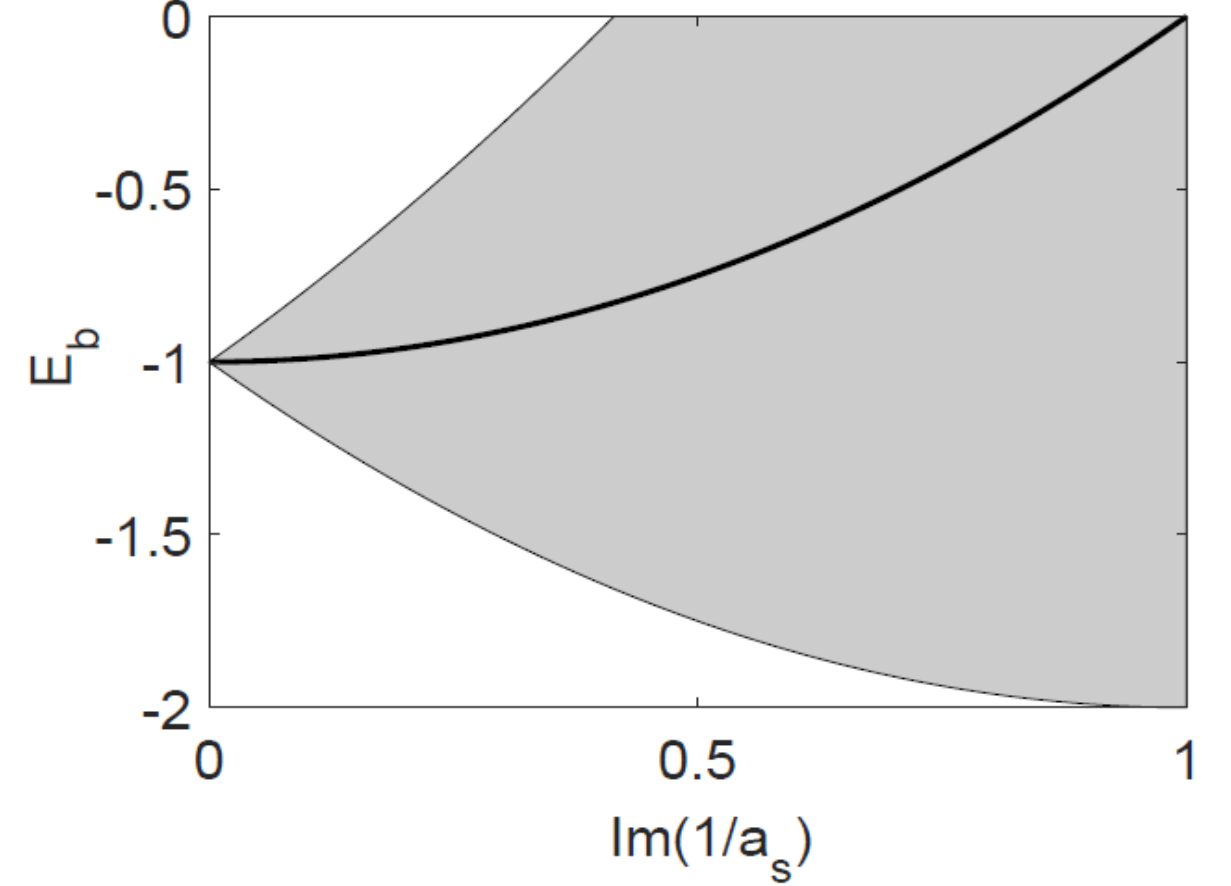}
\caption{ Two-body binding energy $E_b$ as a function of ${\rm Im}(1/a_s)$. Here the black line shows ${\rm Re}(E_b)$, denoting the spectral peak in the molecule detection; the gray area marks the region shifted from   the line by $\pm|{\rm Im}(E_b)|$, denoting the spectral width. 
The units of momentum and energy are respectively ${\rm Re}(1/a_s)\ (>0)$ and ${\rm Re}^2(1/a_s)/m$. } \label{2_body}
\end{figure}

 Experimentally, the binding energy $E_b$ can be detected through the spectroscopy measurement, where the real part ${\rm Re}(E_b)$ determines the position of spectral peak  and the imaginary part ${\rm Im}(E_b)$ determines the spectral width. In Fig.\hspace{1.5mm}\ref{2_body}, we plot out $E_b$ as a function of ${\rm Im}(1/a_s)$, where the black line shows ${\rm Re}(E_b)$ and the gray area marks the energy region shifted from ${\rm Re}(E_b)$ by  $\pm|{\rm Im}(E_b)|$. It shows that for a given ${\rm Re}(1/a_s)>0$, as increasing ${\rm Im}(1/a_s)$ the real part $|{\rm Re}(E_b)|$ will be gradually reduced, until touching zero at ${\rm Im}(1/a_s)={\rm Re}(1/a_s)$. In this process, the imaginary part $|{\rm Im}(E_b)|$ also increases continuously, implying the lifetime of the bound sate becomes gradually shorter and the spectral width becomes broader. In particular, at ${\rm Im}(1/a_s)=0.414{\rm Re}(1/a_s)$ we have $|{\rm Im}(E_b)|=|{\rm Re}(E_b)|$, i.e., the width and the peak position of the spectrum are comparable. Therefore, for  ${\rm Im}(1/a_s)\gtrsim 0.414{\rm Re}(1/a_s)$, the two-body bound state is no longer well defined, as it does not have clear signature in realistic detection.

\section{Efimov physics}\label{sec_IV}

To study the Efimov bound state, we adopt two well established approaches, namely, the Born-Oppenheimer approximation for heavy-heavy-light system and the exact hyper-spherical coordinate approach for three identical bosons. We will show that the two approaches give consistent results regarding the effect of two-body dissipation on Efimov physics.

\subsection{Born-Oppenheimer approximation}

Here we consider  a three-body system where a light atom with mass $m$ interacts with two heavy atoms with mass $M\ (\gg m)$. According to the Born-Oppenheimer approximation (BOA), we first assume the heavy atoms located at fixed positions $-{\cp R}/2$ and ${\cp R}/2$ and write down the wave function of the light atom as
\begin{align}
\psi_{\cp{R}}(\cp{r})\propto
C_1\frac{e^{-\kappa(R)|\cp{r}-\cp{R}/2|}}{|\cp{r}-\cp{R}/2|}+C_2
\frac{e^{-\kappa(R)|\cp{r}+\cp{R}/2|}}{|\cp{r}+\cp{R}/2|},
\end{align}
with energy $\epsilon(R)=-\kappa^2(R)/(2m)$ (here $R\equiv |{\cp R}|)$. Applying the Bethe-Peierls boundary condition to $\psi_{\cp{R}}(\cp{r})$ at $\cp{r}\rightarrow\pm\cp{R}/2$, one obtains two equations for $C_1/C_2$ and $\kappa(R)$. The larger $\kappa(R)$ (corresponding to the lower energy solution) is associated with $C_1/C_2=1$ and given by
\begin{align}
\kappa(R) - e^{-\kappa(R)R}/R=\frac{1}{a_s}. \label{kappa}
\end{align}
In the absence of two-body dissipation and at resonance, the above equation gives $\kappa(R)=0.567/R$ and thus the moving of the light atom results in a scale-invariant potential for the heavy atoms: $\epsilon(R)=-0.161/(mR^2)$. Such scale-invariant potential can result in a sequence of Efimov bound states with discrete scaling symmetry. When the two-body dissipation is present, i.e., the right-hand-side of Eq.(\ref{kappa}) is non-zero ($=i{\rm Im}(1/a_s))$ even at ${\rm Re}(1/a_s)=0$. In the perturbative regime ${\rm Im}(1/a_s) R\ll 1$, we find that the induced interaction 
$\epsilon(R)$ can be expanded as
\begin{align}
\epsilon(R)= -\frac{0.161}{mR^2}-i\frac{0.362}{mR}{\rm Im}(1/a_s) + ... \label{Coulomb}
\end{align}
It tells that the two-body dissipation can generate an imaginary Coulomb potential on top of the $-1/R^2$ potential at short distance, which is expected to destroy the discrete scaling symmetry of Efimov bound states.

\begin{figure}[t]
\includegraphics[width=8cm]{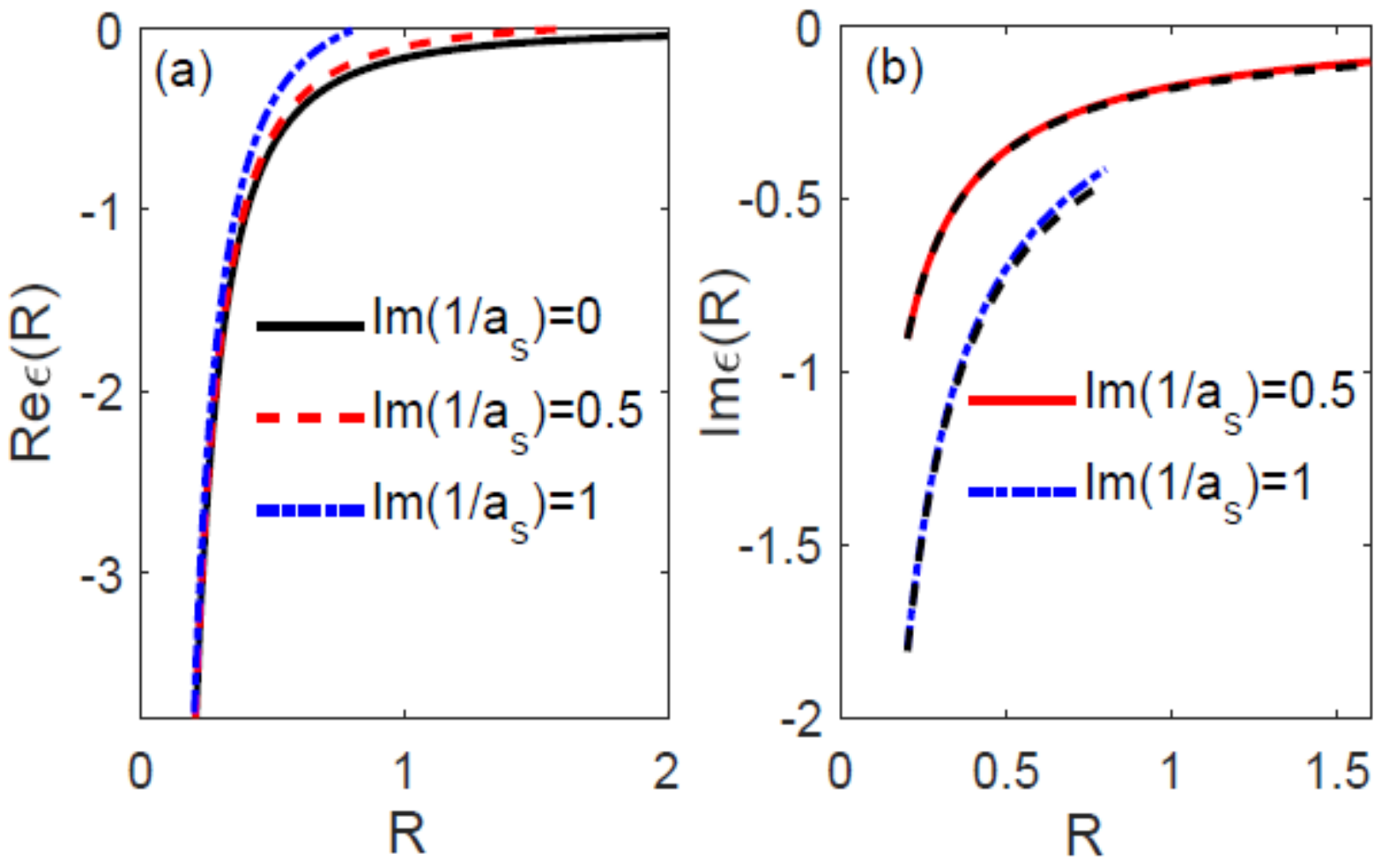}
\caption{(Color online)  Real(a) and imaginary(b) parts of the induced interaction $\epsilon(R)$ under Born-Oppenheimer approximation at ${\rm Re}(1/a_s)=0$ and different ${\rm Im}(1/a_s)=0$. The dashed lines in (b) show the fit to imaginary Coulomb potential as shown in Eq.(\ref{Coulomb}).  Here we take the length unit as $l_0=200 R_c$, where $R_c$ is the short-range cutoff of the wave function, and the energy unit is $E_0=1/(ml_0^2)$. 
} \label{fig_epsilon}
\end{figure}

As the second step in BOA, we solve the Schr{\"o}dinger equation for the heavy atoms:
\begin{align}
[-\frac{\partial^2}{\partial R^2}-\frac{2}{R}\frac{\partial}{\partial R}+\frac{l(l+1)}{R^2}+M(\epsilon(R)-E)]\phi(R)=0. \label{final_eq}
\end{align}
Without two-body dissipation, this equation can be reduced to $[-\frac{\partial^2}{\partial R^2}+\frac{\beta}{R^2}]\chi(R)=0$, with $\chi(R)=R\phi(R)$ and $\beta=l(l+1)-0.16M/m$. When $\beta<-1/4$, Eq.(\ref{final_eq}) results in an infinitely many Efimov bound states with discrete scaling symmetry: $E_{n+1}/E_n=e^{-2\pi/s_0}$, where $s_0=\sqrt{-\beta-1/4}$ and $E_{n}$ is the energy of the $n$-th lowest bound state.

When turn on the two-body dissipation, $\epsilon(R)$ is no longer scale-invariant and Eq.(\ref{final_eq}) cannot be analytically solved. In this case we have numerically solved Eqs.(\ref{kappa},\ref{final_eq}) to obtain the bound state solutions.
As here we have not considered the  effective range term in the scattering length expansion, which corresponds to a zero-range model, the energy of three-body system is not lower bounded. In order to pin down the spectrum in our numerics, we have set a short-range cutoff at $R=R_c$ with $\phi(R_c)=0$, and we take the length unit as $l_{0}= 200R_c$ and the energy unit as $E_0=1/(ml_0^2)$. We have considered the Li-Cs-Cs combination with $M/m=133/6$ and the relative angular momentum between two heavy atoms is zero.

In Fig.\hspace{1.5mm}\ref{fig_epsilon}, we have plotted the induced $\epsilon(R)$ at ${\rm Re}(1/a_s)=0$ for several different dissipation strengths by solving Eq.(\ref{kappa}). Indeed at short distance, the real part of $\epsilon(R)$, ${\rm Re}(\epsilon)$, does not change visibly with different ${\rm Im}(1/a_s)$, and still follows the $-1/R^2$ form. However, the imaginary part of $\epsilon(R)$ sensitively depends on the values of  ${\rm Im}(1/a_s)$, which can be well fit by the Coulomb form as in Eq.(\ref{Coulomb}).
 At large distance, we have $\epsilon(R\rightarrow\infty)\rightarrow E_{b}$, i.e., the two-body binding energy. For   ${\rm Re}(1/a_s)=0$ and at a finite ${\rm Im}(1/a_s)$,  we have ${\rm Re}(E_b)>0$, and therefore ${\rm Re}(\epsilon(R))$ can cross zero at certain $R$ (Fig.\hspace{1.5mm}\ref{fig_epsilon}(a)) and finally saturate at a positive value at $R\rightarrow\infty$.

\begin{figure}[t]
\includegraphics[width=8cm]{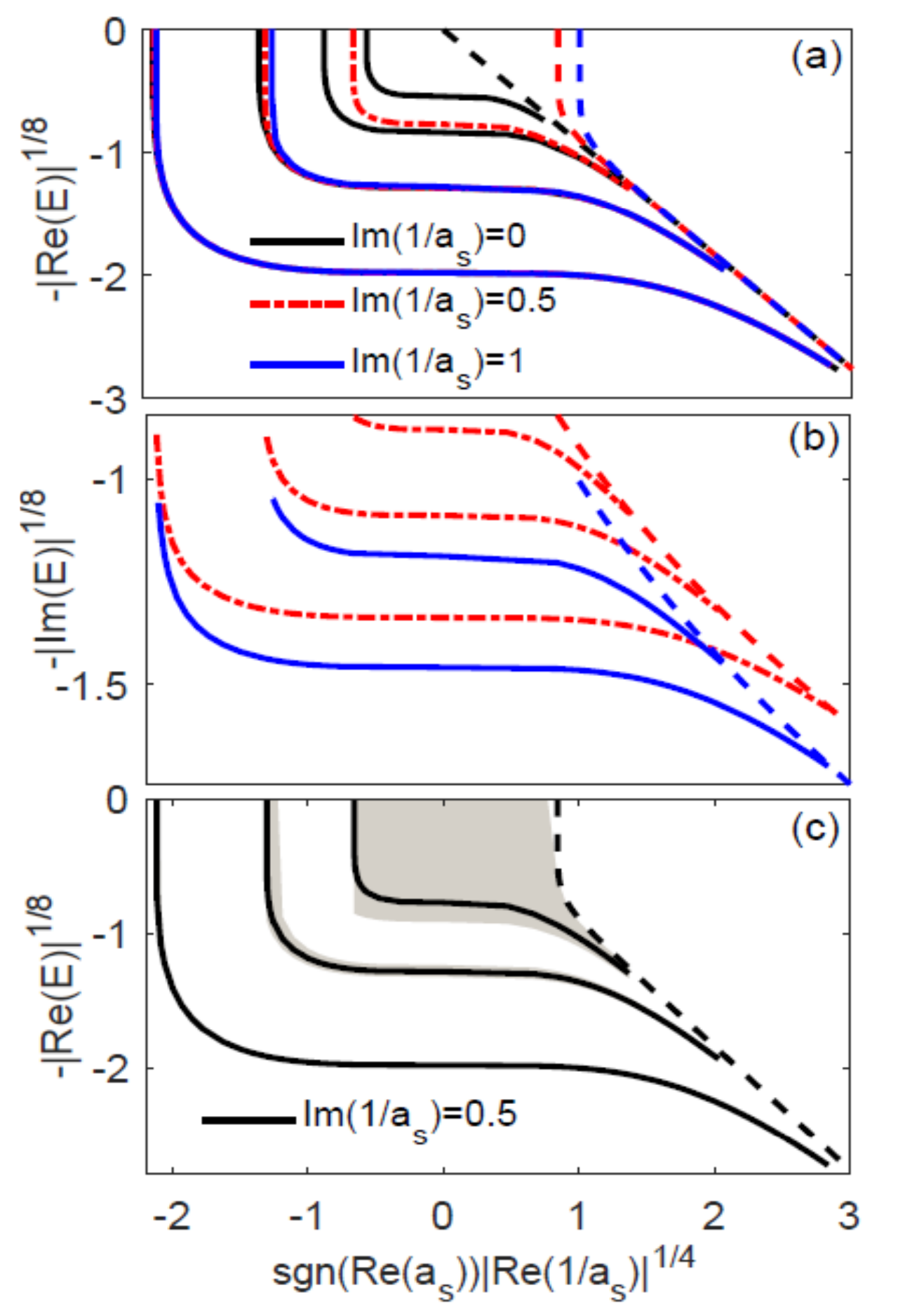}
\caption{(Color online)  Real(a) and imaginary(b) parts of three-body spectra (solid lines) with different two-body dissipation strengths. The dashed lines show the atom-dimer threshold.  (c) Three-body spectrum with a fixed ${\rm Im}(1/a_s)=0.5$. Each black line denotes ${\rm Re}(E)$ and each gray area mark the energy window shifted from the line by $\pm |{\rm Im}(E)|$. Here the mass ratio is $M/m=133/6$, and the units of length and energy are respectively $l_0$ and $E_0=1/(ml_0^2)$. 
 } \label{fig_efimov}
\end{figure}

In Fig.\hspace{1.5mm}\ref{fig_efimov}(a,b), we show the real and imaginary parts of typical three-body spectra without and with two-body dissipations, which respectively correspond to zero and finite ${\rm Im}(1/a_s)$. One can see that the two-body dissipation can significantly influence the shallow bound states by reducing their binding energies, but has little effect on deep ones. Clearly, the discrete Efimov scaling rule for the bound states no longer holds for a finite ${\rm Im}(1/a_s)$, since the shallow bound states can even vanish.  In Fig.\hspace{1.5mm}\ref{fig_efimov}(c), we plot the real and imaginary parts of three-body spectrum in a single figure for a given finite ${\rm Im}(1/a_s)$. Indeed, it is found that for shallow bound states, the amplitude of ${\rm Im}(E)$ (denoting the spectral width) can be larger than that of ${\rm Re}(E)$ (denoting the peak position of the spectrum), and therefore those bound states cannot be well resolved in realistic detection. This is in contrast to deep bound states, whose spectral widths are much smaller than their individual peak positions or the spectral distance between two adjacent levels. These states are much less affected by the two-body dissipation.

\begin{figure}[t]
\includegraphics[width=8cm]{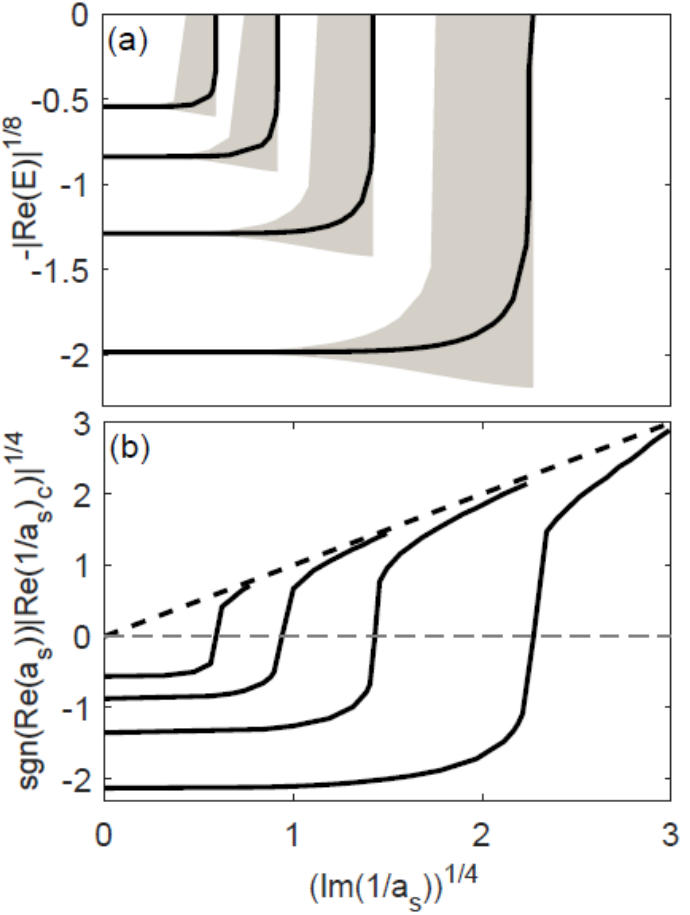}
\caption{(a) Three-body spectrum at ${\rm Re}(1/a_s)=0$ as a function of ${\rm Im}(1/a_s)$.  Each black line denotes ${\rm Re}(E)$ and each gray area mark the energy window shifted from the line by $\pm |{\rm Im}(E)|$.
(b) Critical ${\rm Re}(1/a_s)_c$ to support a three-body bound state as a function of  ${\rm Im}(1/a_s)$, The black dashed line in (b) represents the dimer threshold line ${\rm Re}(1/a_s)={\rm Im}(1/a_s)$.  The intersections between these curves and any radial line from the origin satisfy the generalized scaling symmetry, see text.   The mass ratio and the length and energy units are the same as in Fig.\hspace{1.5mm}\ref{fig_efimov}.} \label{fig_scale}
\end{figure}

To highlight the effect of ${\rm Im}(1/a_s)$ to individual bound states, we examine two characteristic quantities in Efimov physics. One is the binding energy ${\rm Re}(E)$ at ${\rm Re}(1/a_s)=0$;  the other is the critical ${\rm Re}(1/a_s)_c$ to support a trimer bound state with ${\rm Re}(E)<0$. We have plotted these two quantities as functions of ${\rm Im}(1/a_s)$ in Fig.\hspace{1.5mm}\ref{fig_scale}(a) and (b). From Fig.\hspace{1.5mm}\ref{fig_scale}(a), we see that as increasing ${\rm Im}(1/a_s)$, the original Efimov trimer states gradually merges into the continuum with growing spectral width, signifying the vanishing of these bound states. Similarly, Fig.\hspace{1.5mm}\ref{fig_scale}(b) shows that the critical ${\rm Re}(1/a_s)_c$ also increases to touch zero and finally merges into the dimer continuum at ${\rm Re}(1/a_s)_c={\rm Im}(1/a_s)$. All these figures show that the two-body dissipation disfavors the formation of trimer bound states, and at sufficiently strong two-body dissipation, these trimers will vanish one by one.

Although the discrete Efimov scaling symmetry breaks down for a given finite ${\rm Im}(1/a_s)$, Fig.\hspace{1.5mm}\ref{fig_scale} implies a generalized scaling rule when incorporating the scaling of ${\rm Im}(1/a_s)$. Namely, if one scales the real and imaginary parts of $1/a_s$ simultaneously as
\begin{equation}
{\rm Re}(\frac{1}{a_s})\rightarrow\lambda {\rm Re}(\frac{1}{a_s}),\ \ \ \ {\rm Im}(\frac{1}{a_s})\rightarrow\lambda {\rm Im}(\frac{1}{a_s}), \label{new_scale}
\end{equation}
where $\lambda=e^{\pi/s_0}$ is the scaling factor, then the  wave function $\phi(R/\lambda)$ still satisfies the Schr\"{o}dinger equation Eq.(\ref{final_eq}) with  energy $E\rightarrow \lambda^{2}E$, and the discrete Efimov scaling symmetry recovers. One can check this discrete symmetry by following any radial line in Fig.\hspace{1.5mm}\ref{fig_scale}(a,b) and comparing the nearby intersections between the radial line and the curves. For instance, in Fig.\hspace{1.5mm}\ref{fig_scale}(b) the values of ${\rm Im}(1/a_s)$ for the intersections between the horizontal line ${\rm Re}(1/a_s)_c=0$ (see gray dashed line) and these curves obey:
\begin{align}
\frac{{\rm Im}(1/a_s)_{n+1}}{{\rm Im}(1/a_s)_{n}}=\frac{1}{\lambda}. \label{scale}
\end{align}
Here the index $n$ refers to the $n$-th lowest bound state.

\begin{figure}[t]
\includegraphics[width=8cm]{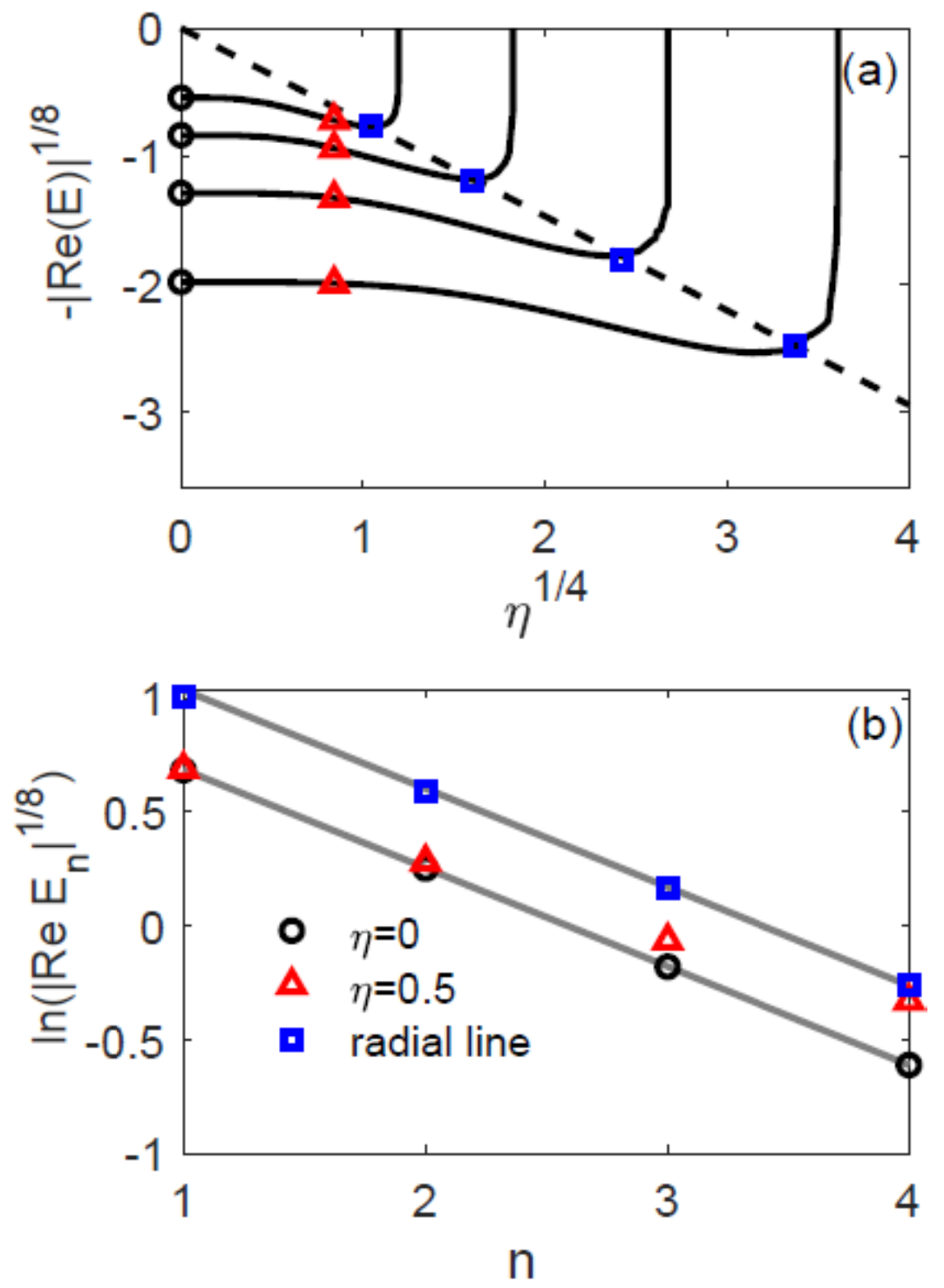}
\caption{(Color online)  (a) Real part of three-body spectrum as a function of  ${\rm Re}(1/a_s)={\rm Im}(1/a_s)\equiv \eta$, when the two-body bound state just emerges from the scattering threshold. (b) Test of discrete scaling symmetry for $\eta=0$ (black circles), $\eta=0.5$ (red triangles), and for varying $\eta$ along the radial line in (a) (blue squares). ``$n$" is the level index. The slope of gray lines is given by $-\pi/(4s_0)$ with $s_0=1.816$, as predicted by the discrete scaling law: $E_{n+1}/E_n=e^{-2\pi/s_0}$. Clearly, the discrete scaling symmetry breaks down for energy levels at $\eta=0.5$. The mass ratio and the length and energy units are the same as in Fig.\hspace{1.5mm}\ref{fig_efimov}.} \label{fig_scale2}
\end{figure}

 Another important question is whether the discrete scaling symmetry can still persist at the emergence of two-body bound state at ${\rm Re}(1/a_s)={\rm Im}(1/a_s)$. This is true in the absence of two-body dissipation (${\rm Im}(1/a_s)=0$), which tells that an infinitely many trimer bound states at two-body resonance follow an intriguing discrete scaling law. However, in the presence of two-body dissipation (${\rm Im}(1/a_s)\neq0$), this is no longer true. To see it, we have plotted out the real part of the spectrum as a function of ${\rm Re}(1/a_s)={\rm Im}(1/a_s)\equiv \eta$ in Fig.\hspace{1.5mm}\ref{fig_scale2}(a), and have checked the discrete scaling symmetry for different $\eta$ in  Fig.\hspace{1.5mm}\ref{fig_scale2}(b). It is found that the discrete scaling symmetry, which predicts $\ln (|E_{n+1}|) = \ln (|E_{n}|) - 2\pi/s_0$,  holds true for $\eta=0$(black circles) but not for a given finite $\eta$ (red triangles). However, if we choose a set of $\eta$ values along a radial line in Fig.\hspace{1.5mm}\ref{fig_scale2}(a), see blue squares for instance, the discrete scaling symmetry will be recovered. These results show that in the presence of two-body dissipation, the discrete scaling law breaks down for a given ${\rm Re}(1/a_s)={\rm Im}(1/a_s)\neq 0$ at the emergence of two-body bound state. Instead, this law can only be recovered by simultaneously change the real and imaginary parts of $1/a_s$ as the rule presented in Eq.(\ref{new_scale}).

\subsection{Hyper-spherical coordinate approach}

Having demonstrated the effect of two-body dissipation to Efimov physics with large mass imbalance, now we switch to the system of three identical bosons using the hyper-spherical coordinate method\cite{Braaten}. Here two essential hyper-spherical coordinates are the hyper-radius  $R$ and hyper-angle $\alpha$, which describe the relative distance of three bosons. According to the low-energy Faddeev equation, the three-body wave function obeys a hyper-spherical expansion\cite{Braaten}:
\begin{align}
\psi(R,\cp{\alpha})=\frac{1}{R^{5/2}\sin{2\alpha}}\sum_n f_n(R)\phi_n(R,\alpha).
\end{align}
Neglecting the off-diagonal couplings between different $n$-levels, $f_n(R)$ follows
\begin{align}
[\frac{1}{2m}(-\frac{\partial^2}{\partial R^2}+\frac{15}{4R^2})+V_n(R)]f_n(R)=Ef_n(R), \label{hyper_H}
\end{align}
with the  hyper-spherical potential
\begin{align}
V_n(R)=[\lambda_n(R)-4]\frac{1}{2mR^2},
\end{align}
where
\begin{align}
\lambda_n^{1/2}\cos(\lambda_n^{1/2}\frac{\pi}{2})-\frac{8}{
\sqrt{3}}\sin(\lambda_n^{1/2}\frac{\pi}{6})=\sqrt{2}\sin(\lambda_n^{1/2}\frac{\pi}{2})\frac{R}{a_s}.  \label{hyper_solution}
\end{align}
At resonance $1/a_s=0$, the lowest eigenvalue $\lambda_0=-s_0^2$ with $s_0=1.00624$. The hyper-spherical potential $V_0(R)$ is purely attractive and scale invariant $\sim -1/R^2$. Then the Schr{\"o}dinger equation Eq.(\ref{hyper_H}) can support an infinitely many bound states with discrete scaling symmetry $E_{n+1}/E_n=e^{-2\pi/s_0}$, where $E_n$ is the $n$-th lowest bound states.

In the presence of two-body dissipation, at ${\rm Re}(1/a_s)=0$, the finite ${\rm Im}(1/a_s)$ will destroy the scaling symmetry. Considering  Eq.(\ref{hyper_solution}) at short distance $R\rightarrow 0$, the lowest eigenvalue can be expanded as
\begin{align}
\lambda_0(R)\approx -s_0^2[1+ i1.897 {\rm Im}(\frac{1}{a_s})R],
\end{align}
which gives
\begin{align}
V_0(R)=-\frac{s_0^2+4}{2mR^2}-i\frac{1.897s_0^2}{2mR} {\rm Im}(\frac{1}{a_s})+ ... \label{V_asymptotic}
\end{align}
It shows that the two-body dissipation contributes an imaginary Coulomb potential to $V_0(R)$. This is consistent with the conclusion from BOA for three-body system with large mass imbalance, see Eq.(\ref{Coulomb}). Similarly, the discrete scaling symmetry for Efimov physics will be recovered by incorporating the scaling of ${\rm Im}(1/a_s)$, as shown in Eq.(\ref{new_scale}).

\begin{figure}[t]
\includegraphics[width=8cm]{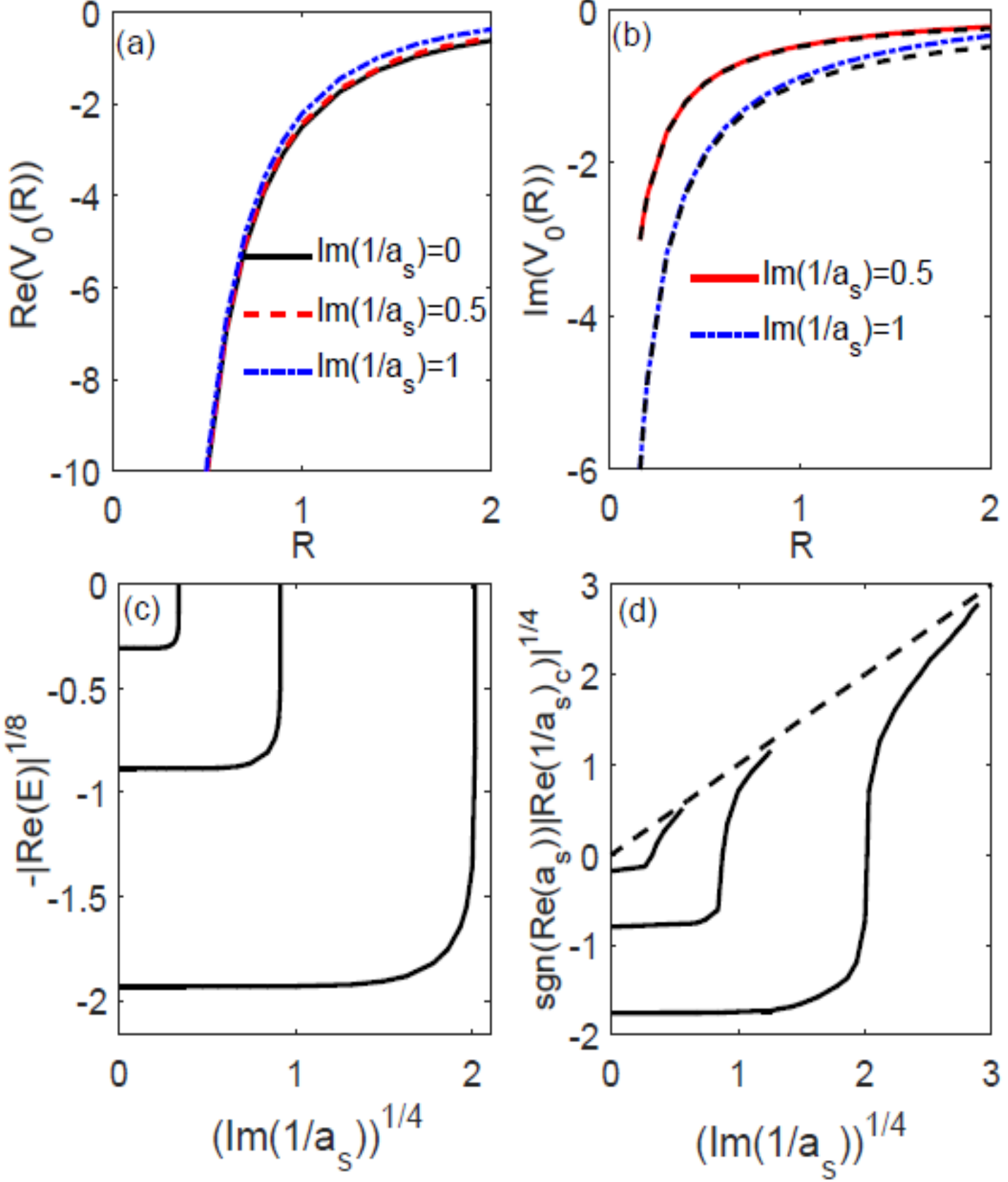}
\caption{(Color online) (a,b) Real and imaginary parts of hyper-spherical potential $V_0(R)$ at  ${\rm Re}(1/a_s)=0$. The dashed lines in (b) show the fit to imaginary Coulomb potential as shown in Eq.(\ref{V_asymptotic}). (c) Real part of three-body spectrum at ${\rm Re}(1/a_s)=0$ as a function of ${\rm Im}(1/a_s)$. (d) Critical ${\rm Re}(1/a_s)_c$ to support a three-body bound state as a function of  ${\rm Im}(1/a_s)$, and the dashed line marks the dimer threshold. The units of length and energy are respectively $l_0$ and $E_0=1/(ml_0^2)$.} \label{fig_hyper}
\end{figure}

Our numerical results  by exactly solving Eqs.(\ref{hyper_solution},\ref{hyper_H}) are shown in Fig.\hspace{1.5mm}\ref{fig_hyper}. Fixing ${\rm Re}(1/a_s)=0$, we show the real and imaginary parts of hyper-spherical potential $V_0(R)$ in Fig.\hspace{1.5mm}\ref{fig_hyper}(a,b). One can see that the effect of finite ${\rm Im}(1/a_s)$ will mainly change the imaginary part of $V_0(R)$ at short distance, which nicely fits the asymptotic behavior in Eq.(\ref{V_asymptotic}). Such change will suppress the bound state formation.  As shown in Fig.\hspace{1.5mm}\ref{fig_hyper}(c) and (d), as increasing ${\rm Im}(1/a_s)$, the binding energy at ${\rm Re}(1/a_s)=0$ will be continuously reduced until zero, and the critical ${\rm Re}(1/a_s)$ to support bound state will increase and finally vanish at the dimer threshold. These results are all consistent with those for the system with large mass imbalance (see Figs.\hspace{1.5mm}(\ref{fig_epsilon},\ref{fig_scale})).

\section{Summary and discussion} \label{sec_V}

In this work, we have studied the effective scattering and Efimov physics with a complex interaction that is induced by two-body dissipation. By employing a two-channel model including the decay of closed-channel molecule, we have shown that the real and imaginary parts of $1/a_s$ can be independently tuned by the detuning and decay rate of closed channel molecules. Importantly, the imaginary part of $1/a_s$, i.e., ${\rm Im}(1/a_s)$, is found to be proportional to the product of  the two-body dissipation strength and the effective range. We then study the effect of two-body dissipation, or the finite ${\rm Im}(1/a_s)$, to the bound state formation in two- and three-body systems. In particular, we have shown that the finite ${\rm Im}(1/a_s)$ can induce an additional imaginary Coulomb potential when three particles come close to each other, thereby suppressing the formation of Efimov bound states and destroying the original Efimov scaling symmetry. Nevertheless, by incorporating the scaling of ${\rm Im}(1/a_s)$ together with that of ${\rm Re}(1/a_s)$, the discrete scaling symmetry can be recovered. We have demonstrated these results for two different three-body systems using the Born-Oppenheimer approximation and the hyper-spherical coordinate approach.

Here we have shown that the non-Hermiticity induced by the two-body dissipation, or the complex $1/a_s$, can suppress the bound state formation in both two-body and three-body systems.   This is in distinct contrast with the effect of other types of non-Hermiticity, for instance, an imaginary magnetic field\cite{non_hermitian_soc1} or non-Hermitian spin-orbit coupling\cite{non_hermitian_soc2} applied in the single-particle level, which are found to facilitate the bound state formation. The comparison suggests  that the effect of  non-Hermiticity on the bound state formation indeed depends on the specific type of non-Hermitian potentials, which can have different origins from one-body or two-body dissipation processes. In this way, the non-Hermiticity can serve as an efficient tool to tune the interaction effect, and a diverse few- and many-body phenomena may emerge from various non-Hermitian systems.

\bigskip

{\bf Acknowledgement.} The work is supported by the National Key Research and Development Program of China (2018YFA0307600), the National Natural Science Foundation of China (No.12074419), and the Strategic Priority Research Program of Chinese Academy of Sciences (No. XDB33000000).

\end{document}